\documentclass[twocolumn,fp]{jpsj3}

\usepackage{dcolumn}
\usepackage{bm}
\usepackage{color}
\usepackage{here}
\usepackage{txfonts}
\def\d{d}

\def\vector#1{{\boldsymbol{#1}}}
\def\va{{\vector a}}
\def\vb{{\vector b}}
\def\vc{{\vector c}}

\def\vH{{\vector H}}
\def\vk{{\vector k}}

\def\vq{{\vector q}}

\def\vr{{\vector r}}

\def\vx{{\vector x}}

\def\vF{{v_{\rm F}}}

\def\pF{{p_{\rm F}}}
\def\Tc{{T_{\rm c}}}

\def\dps{\displaystyle}

\def\kB{{k_{\rm B}}}

\def\TMTSFX{\mbox{${\rm (TMTSF)_2}X$}}

\def\hsp#1{\hspace{#1ex}}

\def\lsim{\stackrel{{\textstyle<}}{\raisebox{-.75ex}{$\sim$}}}
\def\gsim{\stackrel{{\textstyle>}}{\raisebox{-.75ex}{$\sim$}}}

\def\eq.#1{Eq.~(\ref{#1})}
\def\eqs.#1{Eqs.~(\ref{#1})}

\def\Hc2{{H_{\rm c2}}}
\def\difHc2{{H'_{\rm c2}}}
\def\dif2Hc2{{H''_{\rm c2}}}

\def\tb{t_b}
\def\ta{t_a}

\def\Ttri{T^{*}}
\def\Htri{H^{*}}

\newcommand\Equation[2]{
\begin{equation}\label{#1} 
#2
\end{equation}
}

\title{
Dimensional Crossover of the Fulde--Ferrell--Larkin--Ovchinnikov State \\
in Strongly Pauli--Limited Quasi--One--Dimensional Superconductors 
}

\author{Nobumi Miyawaki and Hiroshi Shimahara}

\inst{
Department of Quantum Matter Science, ADSM, Hiroshima University, 
Higashi-Hiroshima 739-8530, Japan
}

\recdate{October 1, 2013}

\abst{
The Fulde--Ferrell--Larkin--Ovchinnikov (FFLO) state 
is examined in quasi-one-dimensional s-wave and d-wave superconductors 
with particular attention paid to the 
effect of the Fermi-surface anisotropy. 
The upper critical field $\Hc2(T)$ is found to exhibit 
a qualitatively different behavior 
depending on the ratio of the hopping energies $\tb/\ta$ 
and the direction of the FFLO modulation vector $\vq$, 
where $\ta$ and $\tb$ are the intra- and interchain hopping energies, 
respectively. 
In particular, when $\tb/\ta \lsim 0.1$ and $\vq \parallel \va$, 
we find a novel dimensional crossover of $\Hc2(T)$ 
from one dimension to two dimensions, 
where $\va$ is the lattice vector of the most conductive chain. 
Just below the tricritical temperature $\Ttri$, 
the upper critical field $\Hc2(T)$ increases steeply 
as in one-dimensional systems, 
but when the temperature decreases, 
the rate of increase in $\Hc2(T)$ diminishes 
and a shoulder appears. 
Near $T = 0$, 
$\Hc2(T)$ shows a behavior typical of the FFLO state 
in two-dimensional systems, 
i.e., an upturn with a finite field at $T = 0$. 
When the angle between $\vq$ and $\va$ is large, 
the upper critical field curve is convex upward at low temperatures, 
as in three-dimensional systems, 
but the magnitude is much larger than 
that of a three-dimensional isotropic system. 
For $\tb/\ta \gsim 0.15$, 
the upper critical fields exhibit a two-dimensional behavior, 
except for a slight shoulder in the range of $0.2 \gsim \tb/\ta \gsim 0.15$. 
The upper critical field is maximum 
for $\vq \parallel \va$ both for s-wave and d-wave pairings, 
while it is only slightly larger than the Pauli paramagnetic limit 
for $\vq \perp \va$. 
The relevance of the present results to the organic superconductor 
${\rm (TMTSF)_2ClO_4}$ is discussed. 
}

\begin{document}
\sloppy
\maketitle

\section{\label{sec:introduction}
Introduction 
}

The Fulde--Ferrell--Larkin--Ovchinnikov (FFLO) state~\cite{Ful64,Lar64} 
has been studied both experimentally and theoretically~\cite{Cas04,Mat07,Leb08}. 
Following the theoretical prediction of Fulde and Ferrell~\cite{Ful64} 
and Larkin and Ovchinnikov~\cite{Lar64}, 
there has been no convincing experimental evidence of this FFLO state. 
However, recently, it has been suggested 
that it occurs in strongly Pauli-limited clean type-II superconductors, 
such as the heavy-fermion superconductor ${\rm CeCoIn}_5$~\cite{Mat07} 
and the quasi-low-dimensional organic superconductors~\cite{
Leb08,Dup95,Shi94,Shi97a,Shi99,Sin00,Sym01,Lor07,Wri11,
Man00,Mak02,Uji12,Hou02,Shi02,Tan02,Yon08a,Yon08b,Leb11,Cro12,
Miy99,Leb10,Aiz09,Fus12} 
$\kappa$--${\rm (BEDT}$-${\rm TTF)_2 Cu(NCS)_2}$~\cite{Sin00,Sym01,Lor07,Wri11}, 
$\lambda$-${\rm (BETS)_2FeCl_4}$~\cite{Uji12,Hou02,Shi02}, 
$\lambda$-${\rm (BETS)_2GaCl_4}$~\cite{Tan02}, 
and ${\rm (TMTSF)_2ClO_4}$~\cite{Yon08a,Yon08b,Leb11,
Cro12,Miy99,Leb10,Aiz09,Fus12}.

Quasi-low-dimensionality is a common feature of the above-mentioned compounds. 
It stabilizes the FFLO state for two reasons. 
First, the orbital pair-breaking effect 
is suppressed when the magnetic field 
is oriented parallel to the conductive layer. 
Particularly in organics, 
the magnetic field must be precisely aligned 
if an FFLO state is to occur~\cite{Shi97b}. 
Second, the highly anisotropic structure of the Fermi surfaces 
in quasi-low-dimensional systems favors the FFLO state. 
The Cooper pairs of the FFLO state have 
a finite center-of-mass momentum $\vq$, 
which characterizes the spatial modulation of the FFLO state. 
In systems with anisotropic Fermi surfaces, 
there exists the optimum direction of $\vq$ 
for which the upper critical field is maximum. 
If $\vq$ can be oriented in the optimum direction, 
the upper critical field is enhanced. 
We term this the Fermi-surface effect hereafter. 
In anisotropic superconductors, the structure of the gap function 
significantly affects the Fermi-surface effect~\cite{Shi97a,Shi99}.

In the present study, the Fermi-surface effect 
in quasi-one-dimensional (Q1D) s-wave and d-wave superconductors 
is investigated, 
motivated by studies of the Q1D organic superconductors \TMTSFX, 
which may exhibit an FFLO state~\cite{Leb08,Shi94,Dup95,Yon08a,Yon08b,
Leb11,Cro12,Miy99,Leb10,Aiz09,Fus12}, 
where 
${\rm TMTSF}$ stands for tetramethyltetraselenafulvalene, 
and $X = {\rm ClO_4}, {\rm PF_6}$, etc. 
In~${\rm (TMTSF)_2PF_6}$, 
Lee et al. found an upturn of $\Hc2(T)$ at low temperatures, 
that exceeds the Pauli paramagnetic limit $H_{\rm P}$~\cite{Lee97}. 
This behavior is consistent with the FFLO state 
in quasi-low-dimensional systems~\cite{Dup95,Shi94,Shi97a,Bur94,Bul73}, 
if the pairing is spin singlet, 
although spin triplet pairing has been suggested~\cite{Lee02}. 
In ${\rm (TMTSF)_2ClO_4}$, Oh et al. found that $\Hc2$ exceeds $H_{\rm P}$ 
at low temperatures~\cite{Oh04}. 
Recently, Yonezawa et al. discovered a shift 
in the principal axis of the in-plane field-angle 
dependence of the superconducting onset temperature, 
which may be related to the FFLO state. 
In this system, 
a singlet state with line nodes (the so-called d-wave state) is 
considered to be likely.

The field-angle dependence of $\Hc2$ 
mainly arises from the anisotropy of the orbital pair-breaking effect 
in conventional superconductors. 
In an FFLO state, 
the in-plane field-angle dependence is strongly affected by 
the Fermi-surface effect via the vector $\vq$.

Lebed has revealed that, 
for a magnetic field in the direction of the $\vb'$-axis, 
the dimensional crossover in the orbital pair-breaking effect 
from three dimensions to two dimensions induces 
hidden reentrant and FFLO phases~\cite{Leb11}. 
The upper critical field in the direction of the $\vb'$-axis 
is estimated to be about $6~{\rm T}$, 
which is in agreement with the experimental result~\cite{Yon08a,Oh04}. 
Croitoru, et al. have investigated 
the in-plane magnetic-field anisotropy of the FFLO state, 
assuming an elliptic Fermi surface, 
and revealed that the superconducting temperature is maximum 
for a field oriented perpendicular to the FFLO vector~\cite{Cro12}. 
Their calculations 
support the interpretation of the experimental result 
for the field-angle dependence as a realization of the FFLO state 
with $\vq \parallel \vb'$. 
The FFLO states for ${\rm (TMTSF)_2}X$ and ${\rm (TMTSF)_2ClO_4}$ 
have been studied, 
taking both the Pauli paramagnetic and orbital pair-breaking effects 
into account~\cite{Miy99,Leb10}. 
It was found that $\Hc2$ is consistent with the experimental data.

In spite of the above studies, the limits of the pure FFLO state 
have not been clarified in Q1D systems. 
We examine this issue, 
with particular attention paid to the Fermi-surface effect. 
If the Fermi surface is warped, 
the direction of the optimum $\vq$ is nontrivial. 
In fact, that is quite different from 
those conjectured from simple physical considerations 
based on the shape of the Fermi surface in some models~\cite{Shi97a,Shi99}.

In order to examine the Fermi-surface effect, 
the concept of Fermi surface ``nesting'' for the FFLO state 
has been introduced~\cite{Shi94,Shi97a,Shi99}, 
in analogy to those for the charge density wave (CDW) 
and spin density wave (SDW). 
Since the FFLO state is due to Cooper pairs of two electrons with 
$(\vk,\uparrow)$ and $(-\vk + \vq,\downarrow)$, 
its stability is closely related to the extent of the overlap 
of the Fermi surfaces of spin-up and spin-down electrons, 
where the latter Fermi surface is inverted and shifted by $\vq$, 
which is expressed as $\vk \rightarrow - \vk + \vq$.

In one dimension, the upper critical field $\Hc2(T)$ diverges 
in the limit $T \rightarrow 0$~\cite{Mac84,Suz83,Buz83}. 
This result is due to perfect nesting, 
which means that the overlap occurs in a finite area 
on the Fermi surface, 
classified as type (a) in Table~\ref{table:nesting}. 
However, 
for realistic interaction strengths between electrons, 
such one-dimensional (1D) systems should undergo CDW or SDW transitions. 
Therefore, the best candidate is a quasi-two-dimensional (Q2D) system, 
in which the CDW and SDW transitions are suppressed. 
In such systems, the Fermi surfaces touch on one or more lines 
by the transformation $\vk \rightarrow - \vk + \vq$ of the spin-down 
Fermi surface.\cite{Shi94,Shi97a,Shi99} 
This type of nesting results in $\difHc2(0) \ne 0$ and $\Hc2(0) < \infty$ 
and the upturn of $\Hc2(T)$ at low temperatures, 
classified as type (b) in Table~\ref{table:nesting}, 
where $\difHc2(T) \equiv \d \Hc2/\d T$. 
In Q2D systems, the nesting enhances the FFLO state, 
while suppressing the CDW and SDW instabilities.~\cite{Shi94}

In this context, Q2D systems include Q1D systems in which 
the interchain hopping energy $\tb$ 
is large enough to suppress the CDW and SDW transitions. 
Although ${\rm (TMTSF)_2}X$ is called Q1D, 
it should be classified as Q2D 
with respect to the nesting effect of the FFLO state.

On the other hand, in isotropic systems with spherical Fermi surfaces, 
the upper critical field of the FFLO state is only slightly higher than 
the Pauli paramagnetic limit of the BCS state with $\vq = 0$. 
In such systems, $q \equiv |\vq|$ at $T = 0$ 
is larger than $2 h/\vF$, 
which is the distance between the Fermi surfaces of the spin-up 
and spin-down electrons, 
because crossing along a line is a better nesting condition 
than touching at a point. 
This results in $\difHc2(0) = 0$, 
classified as type (c) in Table~\ref{table:nesting}.

The characteristic behaviors of $\Hc2(T)$ at low temperatures 
are summarized in Table~\ref{table:nesting}. 
A Q1D organic superconductor at low temperatures 
should be classified as type (b), because of the warp in the Fermi surface. 
However, at high temperatures, 
the behavior of the upper critical field can be more complicated, 
owing to the shape of the Fermi surface, the density of states, 
and the gap anisotropy. 
In fact, in the intermediate temperature region $T \lsim \Ttri$, 
hybrid behaviors of types (a) to (c) occur, depending on 
$\tb/\ta$ and $\varphi$, 
where $\varphi$ is the angle between $\vq$ and the crystal $\va$-axis.

The paper is organized as follows. 
In~Sect.~\ref{sec:nesting}, the nesting effect of the FFLO state 
in Q1D systems is discussed. 
In~Sect.~\ref{sec:numericalresults}, 
the transition temperature equation and numerical results 
are presented. 
In~Sect.~\ref{sec:last}, the results are summarized and discussed. 
Units in which $\hbar = 1$ and $\kB = 1$ are used throughout. 
For the organic conductors, quarter-filled bands are assumed.

\begin{table}[hbtp]
\begin{center}
\caption{
Nesting conditions and low-temperature behaviors of $\Hc2(T)$. 
}
\label{table:nesting}
\begin{tabular}{c}
\\
\begin{tabular}{l|l|c|c|c}
type  & \multicolumn{1}{c|}{nesting} 
        & $\Hc2(0)$   &  $\difHc2(0)$  &  $\dif2Hc2(T) $  \\[2pt]
\hline 
(a) & 
      touch on a surface 
                & $\infty$    &  N/A           &  $ > 0$          \\
(b) & 
      touch on a line 
                & finite     &  $ < 0$         &  $ > 0$          \\
(c) & 
      crossing along a line 
                & finite     &  0              &  $ < 0$ 
\end{tabular}
\end{tabular}
\end{center}
\end{table}

\section{\label{sec:nesting}
Nesting Effect in Q1D Systems
}

We consider a model having Q1D energy dispersion 
\Equation{eq:Q1Depsilon}
{
     \xi_{\sigma}(\vk,h) 
       = - 2 \ta \cos ( \vk \cdot \va ) 
         - 2 \tb \cos ( \vk \cdot \vb ) 
         - h \sigma - \mu , 
}
with $h = \mu_{\rm e} | \vH |$, 
where $\mu$ and $\mu_{\rm e}$ are the chemical potential 
and the magnitude of the electron magnetic moment, respectively. 
Equation (\ref{eq:Q1Depsilon}) assumes that $t_a > t_b \gg t_c$; 
it omits the interplane hopping energy $t_c$ for simplicity. 
However, it is supposed that $t_c$ is large enough to 
stabilize the superconducting long-range order 
and justify the mean-field approximation, 
but that it is small enough to be omitted 
in the resultant mean-field self-consistent equation. 
Introducing the reciprocal lattice vectors 
${\bar \va}$, ${\bar \vb}$, and ${\bar \vc}$ 
and the momentum components $k_x$, $k_y$, and $k_z$ via 
$\vk = k_x {\bar \va} + k_y {\bar \vb} + k_z {\bar \vc}$ 
leads to 
\Equation{eq:Q1Depsilonpxpy}
{
     \xi_{\sigma}(\vk,h) 
       = - 2 \ta \cos ( k_x ) 
         - 2 \tb \cos ( k_y )
         - h \sigma - \mu . 
}
For simplicity of notation, $k_x$ and $k_y$ have been redefined 
to include the lattice constants $a$ and $b$, respectively. 

In order to discuss the nesting effect, 
we define the energy difference as 
\Equation{eq:Deltaepsilon}
{
     \Delta \epsilon(k_y,\vq,h) 
     \equiv 
     \Bigl [
       \xi_{\uparrow}(\vk) 
     - \xi_{\downarrow}(- \vk + \vq) 
     \Bigr ]_{k_x = k_{{\rm F}x}^{\uparrow}(k_y)} , 
     }
with $k_{{\rm F}x}^{\sigma}(k_y)$ denoting a positive function 
that satisfies 
$
     \xi_{\sigma} ( k_{{\rm F}x}^{\sigma}(k_y), k_y) = 0 
     $. 
On the Fermi surface, we define 
$\Delta k_{{\rm F}x}(k_y,\vq) \equiv 
k_{{\rm F}x}^{\uparrow}(k_y)
- k_{{\rm F}x}^{\downarrow}(k_y - q_y)  - q_x$. 
For the vector $\vq$ that satisfies $\Delta k_{{\rm F}x}(k_y,\vq) = 0$, 
the energy difference 
$
     \Delta \epsilon(k_y,\vq,h) 
$ is zero. 
If there exists a constant vector $\vq$ such that 
$\Delta k_{{\rm F}x}(k_y,\vq) = 0$ over a finite range of $k_y$ values, 
perfect nesting occurs. 
In such a case, $\Hc2(T)$ diverges in the limit $T \rightarrow 0$. 
However, such a constant vector $\vq$ does not exist when $\tb \ne 0$.

The nesting condition is not correctly treated 
by the linearized energy dispersion relation 
$     \xi_{\sigma}^{(\pm)} (\vk,h) 
      \approx {\bar v}_{\rm F} \, (k_x \pm \pF(k_y,k_z) ) 
        + \epsilon_{\perp}(k_y,k_z) - h \sigma - \mu$ 
with a constant Fermi velocity ${\bar v}_{\rm F}$, 
as adopted by many previous authors. 
In this model, 
the displacement of the Fermi surfaces of spin-up and spin-down electrons 
due to the Zeeman energy $2 h$ is compensated for 
by the constant shift $\vq = ({2 h}/{{\bar v}_{\rm F}},0,0)$ 
independent of $(k_y,k_z)$, 
which implies perfect nesting for the FFLO state. 
However, in realistic Q1D systems, 
the Fermi velocity $\vF$ depends on $k_y$ 
with a variation from ${\bar v}_{\rm F}$ on the order of 
$\tb {\bar v}_{\rm F}/\ta$, 
which is small but nonnegligible. 
This variation in $\vF$ causes a mismatch of the Fermi surfaces, 
which significantly affects $\Hc2(T)$ at low temperatures.

Equation (\ref{eq:Q1Depsilonpxpy}) gives the magnitude of 
the Fermi velocity along the most conductive chain as 
$\vF = 2 \ta \sin (k_{{\rm F}x})$ in the unit of $a/\hbar$, 
where $k_{{\rm F}x}$ denotes the Fermi momentum in the chain direction. 
For quarter-filled bands, since $k_{{\rm F}x} \sim \pi/4$, 
one obtains $\vF \approx \sqrt{2} \ta \equiv {\bar v}_{\rm F}$. 
The energy difference $\Delta \epsilon(k_y,\vq,h)$ 
due to $\Delta \vF \equiv \vF - {\bar v}_{\rm F}$ 
is estimated to be 
$\dps{ {\Delta \vF} q  \lsim  \tb \vF q/\ta \sim \tb \Tc^{(0)}/\ta}$, 
since $\vF q \sim h \sim \Tc^{(0)}$, 
using a value of $\vq$ that makes the Fermi surfaces touch on a line, 
where $\Tc^{(0)}$ denotes the zero-field transition temperature. 
Therefore, the crossover temperature $T_0$ between the perfect and imperfect 
nesting conditions is proportional to $\tb \Tc^{(0)}/\ta$.

At higher temperatures $\Ttri \gsim T \gsim T_0$, 
$\Delta \epsilon$ is negligible 
in comparison to the thermal energy $\kB T$. 
Hence, the small mismatch in the Fermi surfaces $\Delta k_{{\rm F}x}$ 
does not affect the upper critical field significantly 
because of the temperature effect. 
Therefore, the system behaves like a 1D system, 
in which the Fermi-surface nesting for the FFLO state is perfect. 
However, at low temperatures, $T \lsim T_0$, 
the variation $\Delta \vF$ due to the warp in the Fermi surface 
can be substantial. 
For $T \sim T_0$, the system begins to lose its 1D character, 
and when $T \ll T_0$, 
the two-dimensional (2D) character of the system is recovered. 
Therefore, when the interchain hopping energy $\tb$ is small enough that 
$\Ttri \gsim T_0 \propto \tb \Tc/\ta$, 
a dimensional crossover between one dimension and two dimensions 
can occur~\cite{Note1}. 
In the next section, 
it is verified by numerical calculations that such a crossover actually occurs.

\section{\label{sec:numericalresults}
Formulation and Numerical Results
}

The extension of the formula for anisotropic FFLO states\cite{Shi97a,Shi99} 
to Q1D systems is straightforward. 
The equation for the transition temperature is 
\Equation{eq:Tceq}
{
     \begin{split}
     \log \frac{\Tc^{(0)}}{\Tc} 
       \, = \, & 
       \int_0^{\infty} \hsp{-0.5} \d t \,\, 
       \sum_{s = \pm}
       \int_{-\pi}^{\pi} 
       \frac{\d k_y}{2 \pi} \frac{\rho_{\alpha}(0,k_y)}{N_{\alpha}(0)} 
     \\
     & \, \, 
     \times \sinh^2 \frac{\beta \zeta}{2} 
       \frac{\tanh t}{t \, ( \cosh^2 t + \sinh^2 (\beta \zeta/2) )} , 
     \end{split}
     }
with 
\Equation{eq:zeta}
{
     \zeta(s,k_y,h,\vq) = \dps{ 
       h \, 
       \Bigl ( 
         \frac{{\vector v}_{\rm F}(s,k_y) \cdot \vq}
              {2h} - 1 
       \Bigr ) 
         } 
     }
and 
\Equation{eq:Nalpha}
{
     N_{\alpha}(0) 
       = 
       \sum_{s = \pm}
       \int_{-\pi}^{\pi} \rho_{\alpha}(0,k_y) \frac{\d k_y}{2 \pi} , 
     }
where we define 
the effective density of states on the Fermi surface 
$     \rho_{\alpha}(0, k_y) 
       = 
       \rho (0,k_y) 
         \bigl [ \gamma_{\alpha} (k_y) \bigr ]^2 
     $. 
Here, $\rho(\xi,k_y)$ is the density of states defined by 
\Equation{eq:rho}
{
     \frac{1}{N} \sum_{\vk} F(\vk)
       = \int \d \xi 
         \sum_{s = \pm} 
         \int_{-\pi}^{\pi} \frac{\d k_y}{2 \pi} 
           \rho(\xi,k_y) F(\xi,s,k_y)
     }
for the arbitrary smooth function 
$F(\xi_{\sigma}(\vk,0),s,k_y) = F(\vk)$ 
with $s = {\rm sgn}(k_x)$. 
The function $\gamma_{\alpha}(k_y)$ expresses the momentum dependence 
of the gap function on the Fermi surface. 
The suffix $\alpha$ denotes the symmetry index. 
Below, we examine two cases: 
$\gamma_{\rm s}(k_y) = 1$ 
and 
$\gamma_{\rm d}(k_y) = \dps{ \sqrt{2} \cos k_y }$. 
These cases are conventionally called 
the s-wave and d-wave states, respectively.

As mentioned above, the momentum dependence of the Fermi velocity 
${\vector v}_{\rm F}(s,k_y)$ is taken into account. 
The FFLO modulation vector $\vq$ is optimized 
so that $\Tc$ or $\Hc2$ is maximized. 
In the presence of the orbital effect (unless it is extremely weak) 
the direction of $\vq$ is locked to the magnetic field direction~\cite{Gru66}. 
In such a situation, 
the magnitude of $\vq$ should be optimized, 
while the direction of $\vq$ is fixed in a given direction.

At $T = 0$, the upper critical field is obtained by solving 
\Equation{eq:hc2T0}
{
     h_{\rm c2} = \frac{\Delta_{\alpha 0}}{2} \exp 
     \left [
       - \sum_{s = \pm} 
         \int_{-\pi}^{\pi} \hsp{-0.5} \frac{\d k_y}{2 \pi} 
           \frac{\rho_{\alpha}(0,k_y)}{N_{\alpha}(0)} 
             \log \Bigl | 
                    1 - \frac{{\vector v}_{\rm F}(s,k_y) \cdot \vq}
                             {2 h_{\rm c2}} 
                  \Bigr | 
     \right ]
     }
with $h_{\rm c2} = \mu_{\rm e} H_{\rm c2}$ and 
$\Delta_{\alpha 0} = 2 \omega_{\rm c} \exp[ - 1/|V_{\alpha}| N_{\alpha}(0) ]$, 
where $\omega_{\rm c}$ and $V_{\alpha}$ are the cutoff frequency 
and the coupling constant of the $\alpha$ component of 
the pairing interactions, respectively.\cite{Shi97a,Shi99}

First, we consider the case in which $\vq \parallel \va$. 
This direction of $\vq$ seems favorable for Fermi-surface nesting, 
because Fermi surfaces touch at the shortest $\vq$, 
and thus the spatial variation in $\Delta(\vr)$ is minimal. 
This can be proven through a detailed analysis that takes into account 
the density of states and the matching of the Fermi surfaces 
displaced by the Zeeman energy, 
as confirmed below by numerical calculations.

\begin{figure}[htbp]
\vspace{2ex} 
\vspace{2ex} 
\begin{center}
\includegraphics[width=8.0cm]{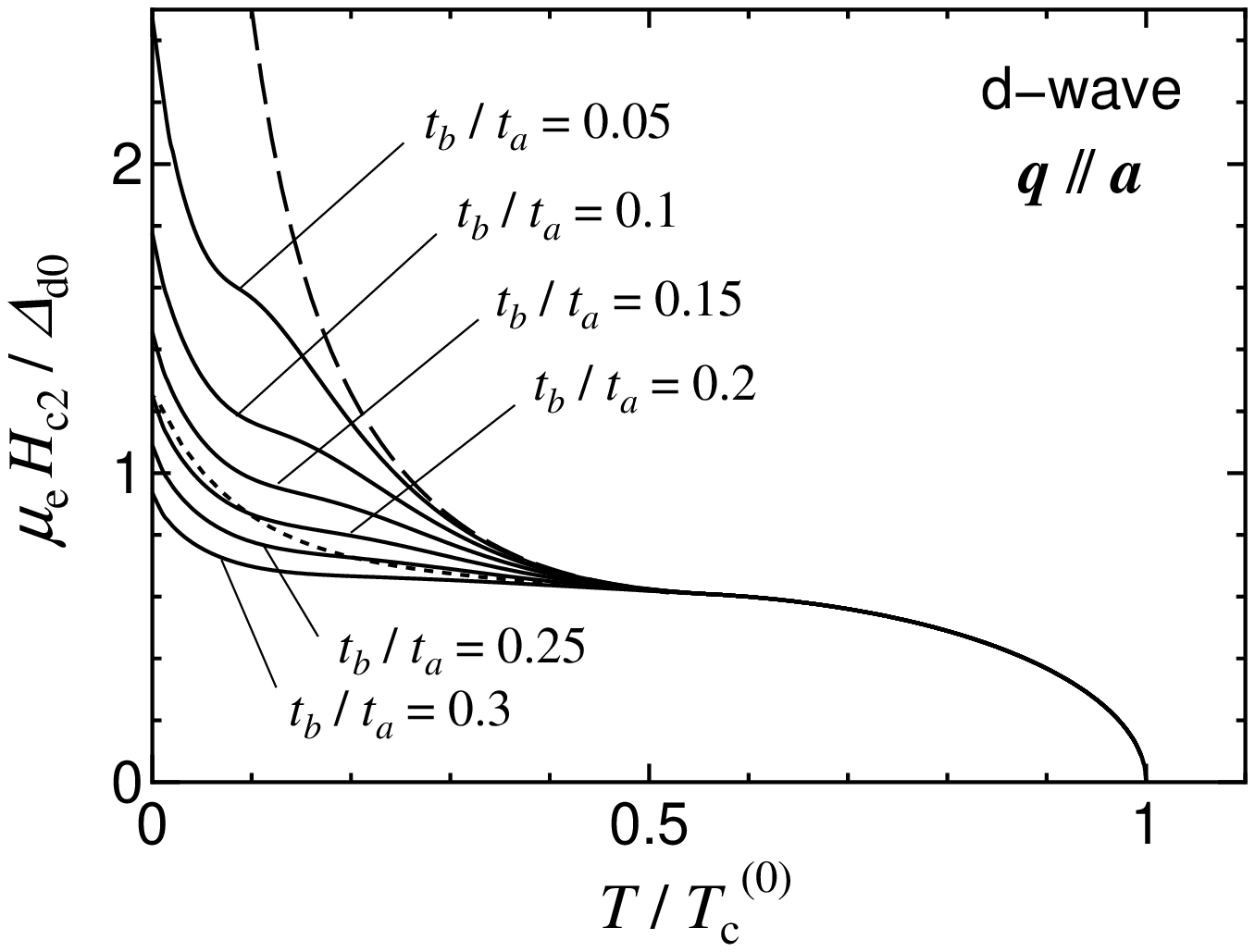}
\end{center}
\caption{Temperature dependence of the upper critical field 
for d-wave pairing, when $\vq \parallel \va$. 
The solid curves show 
the results for the range of $\tb/\ta$ from 0.05 to 0.3. 
The dashed curve is for a 1D system at $\tb = 0$. 
The dotted curve is for a 2D isotropic system 
with ${\rm d}_{x^2-y^2}$-wave pairing when 
$\vq \parallel {\hat \vx}$. 
}
\label{fig:hc2-T_vrtb_phi0}
\end{figure}

Figure~\ref{fig:hc2-T_vrtb_phi0} shows 
the temperature dependence of the upper critical fields for d-wave pairing. 
For $\tb/\ta \lsim 0.1$, a new kind of dimensional crossover arises 
as follows. 
Just below the tricritical temperature $\Ttri$, 
the upper critical field $\Hc2(T)$ increases steeply 
along the curve of the 1D system, 
but as the temperature decreases, the rate of increase in $\Hc2(T)$ is reduced, 
and a shoulder appears. 
At lower temperatures, 
it reduces to the behavior of 2D systems, 
i.e., it exhibits an upturn with a finite value at $T = 0$. 
The shoulder becomes less pronounced for $\tb/\ta \sim 0.15$ 
and completely disappears for $\tb/\ta \sim 0.25$. 
Independently of $\tb/\ta \ne 0$, 
the low-temperature behavior is essentially that of a Q2D system 
and thus classified as type (b). 
As $\tb/\ta$ increases, the FFLO upper critical field decreases, 
and for $\tb/\ta \gsim 0.25$, the upper critical field 
is lower than that in 2D isotropic systems.

\begin{figure}[htbp]
\vspace{2ex} 
\vspace{2ex} 
\begin{center}
\includegraphics[width=8.0cm]{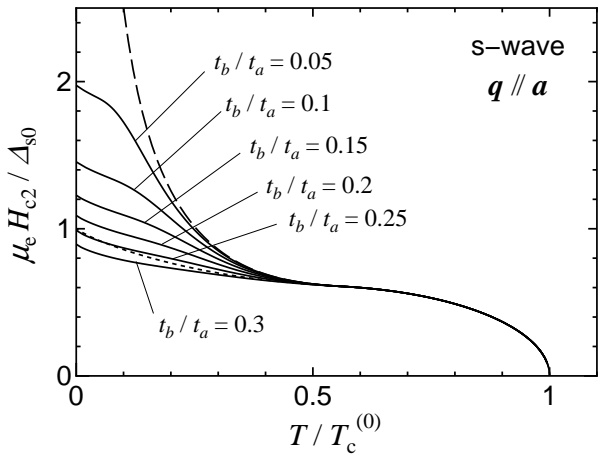}
\end{center}
\caption{Temperature dependence of the upper critical field 
for s-wave pairing when $\vq \parallel \va$. 
The solid curves are for the range of $\tb/\ta$ from 0.05 to 0.3. 
The dashed curve plots a 1D system at $\tb = 0$. 
The dotted curve is for a 2D isotropic system with s-wave pairing. 
}
\label{fig:hc2-T_vrtb_phi0_sw}
\end{figure}

Figure~\ref{fig:hc2-T_vrtb_phi0_sw} shows the results for s-wave pairing. 
Similarly to d-wave pairing, 
a dimensional crossover from one dimension to two dimensions 
is found in the temperature dependence, 
but the upturn at low temperatures is weaker than that for d-wave pairing, 
owing to the difference in the Fermi-surface nesting. 
For the nesting vector $\vq \parallel \va$, 
the Fermi surfaces touch on a line at $k_y = 0$, 
where the amplitude of the gap function is maximum for d-wave pairing.

\begin{figure}[htbp]
\vspace{2ex} 
\vspace{2ex} 
\begin{center}
\includegraphics[width=8.0cm]{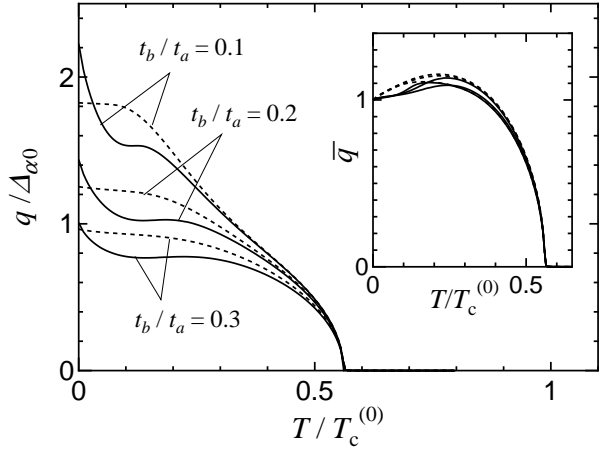}
\end{center}
\caption{
Temperature dependence of $q = |\vq|$ when $\vq \parallel \va$. 
The solid and dashed curves show the results 
for d-wave and s-wave pairings, respectively. 
The inset shows the temperature dependence of 
${\bar q} \equiv v_{\rm F0} q/2h$. 
}
\label{fig:vq}
\end{figure}

Figure~\ref{fig:vq} shows the temperature dependence of 
the optimum $q \equiv |\vq|$ 
along the upper critical field curve $H = \Hc2(T)$, 
when $\vq \parallel \va$. 
Below $T \approx 0.56 \times \Tc$, the FFLO state $\vq \ne 0$ occurs. 
For d-wave pairing, the behavior of $q$ is not monotonic, 
reflecting the behavior of $\Hc2(T)$. 
As shown in the inset, all of the curves of 
${\bar q} \equiv v_{\rm F0} q/2h$ 
converge to unity at $T = 0$, 
where $v_{\rm F0} \equiv |v_{{\rm F}x}(s,k_y=0)|$. 
This convergence implies that the Fermi surfaces touch on a line at $k_y = 0$ 
by the transformation $\vk \rightarrow - \vk + \vq$ of the spin-down 
Fermi surface.

\begin{figure}[htbp]
\vspace{2ex} 
\vspace{2ex} 
\begin{center}
\includegraphics[width=8.0cm]{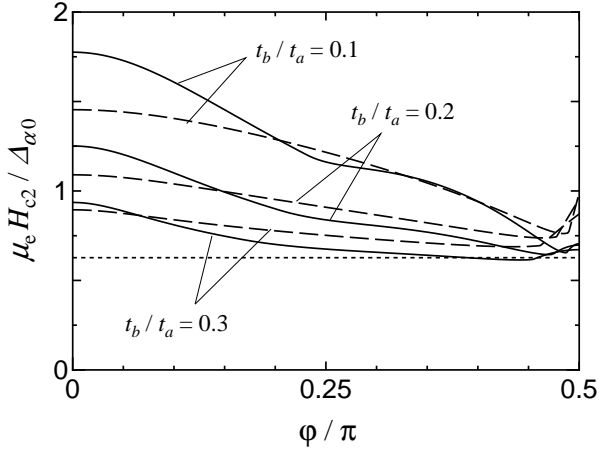}
\end{center}
\caption{Angle $\varphi$ dependence of the upper critical fields 
at $T = 0$. 
The solid and dashed curves show 
the results for d-wave and s-wave pairings, respectively. 
The dotted curve shows the Pauli paramagnetic limit 
for d-wave pairing with $\tb/\ta = 0.3$, 
calculated using the formula in Ref.~\citen{Shi97a}. 
}
\label{fig:hc2-phi_various_tb_dw_sw}
\end{figure}

Next, we consider the dependence of $\Hc2(T)$ 
on the direction of the FFLO modulation vector $\vq$. 
Figure~\ref{fig:hc2-phi_various_tb_dw_sw} shows $\Hc2(0)$ as a function 
of the angle $\varphi$ for d-wave and s-wave pairings. 
The direction $\vq \parallel \va$ is the most favorable 
for the nesting effect at $T = 0$, 
for both d-wave and s-wave pairings. 
This result remains unchanged at finite temperatures.

\begin{figure}[htbp]
\vspace{2ex} 
\vspace{2ex} 
\begin{center}
\includegraphics[width=8.0cm]{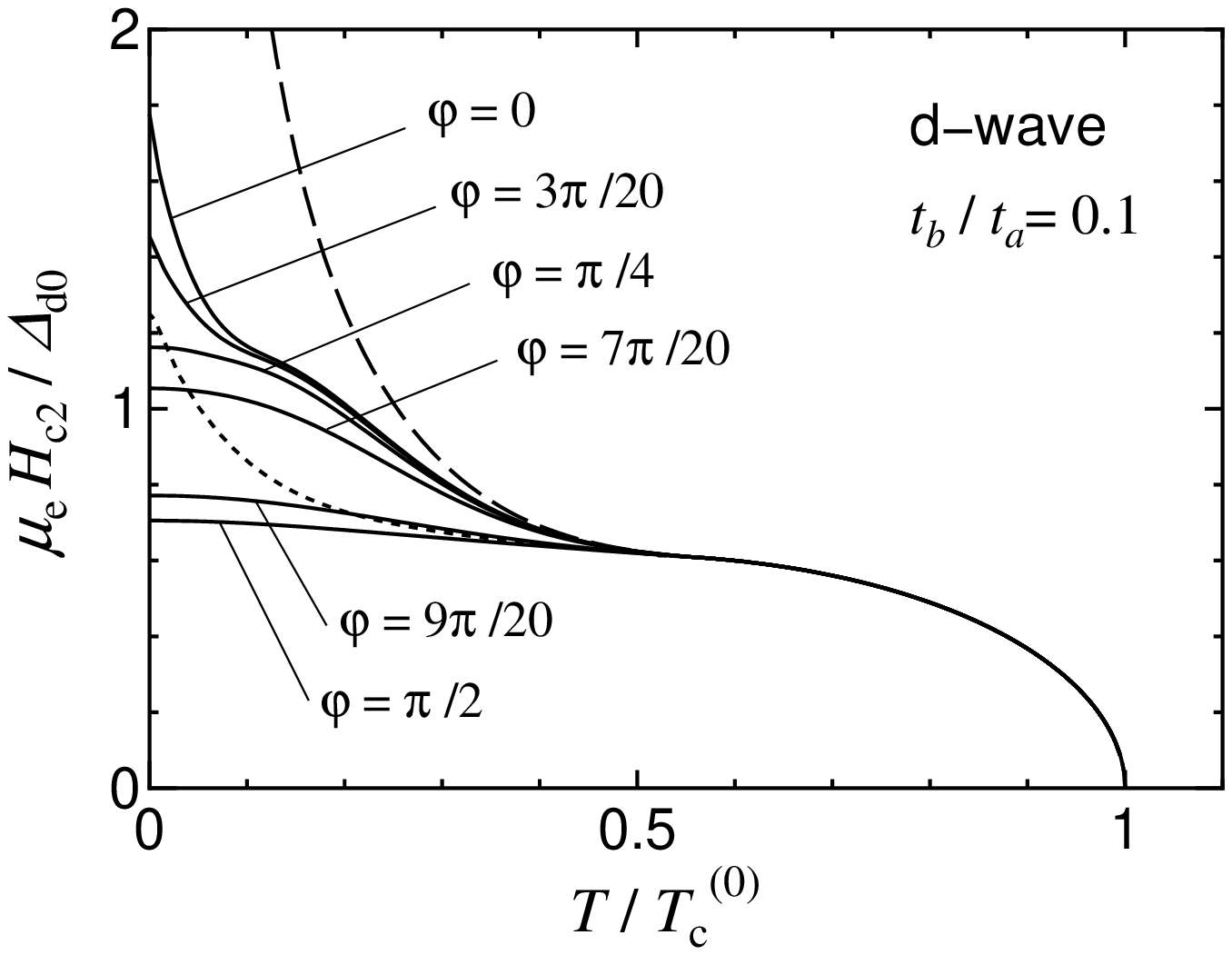}
\end{center}
\caption{
Temperature dependence of the upper critical field 
for d-wave pairing when $\tb/\ta = 0.1$. 
The solid curves show 
the results for the range of $\varphi$ from 0 to $\pi/2$. 
The dashed curve is for a 1D system. 
The dotted curve is for a 2D isotropic system 
with ${\rm d}_{x^2-y^2}$-wave pairing when 
$\vq \parallel {\hat \vx}$. 
}
\label{fig:hc2-T_various_phi_tb010}
\end{figure}

Figure~\ref{fig:hc2-T_various_phi_tb010} 
shows the results for d-wave pairing 
when $\tb/\ta = 0.1$. 
The upturn in the low-temperature region disappears 
for $\varphi \gsim \pi/4$, 
but the large shoulder remains. 
This behavior can be interpreted in terms of the nesting concept. 
For a large angle $\varphi \gsim \pi/4$, 
the nesting condition of crossing along lines 
becomes more effective than that of touching on a line 
near the node of the d-wave gap function. 
Therefore, the low-temperature behavior is classified as type (c), 
but the magnitude is much larger than 
that in three-dimensional (3D) isotropic systems, 
because the crossing angle between the Fermi surfaces is extremely small 
owing to the Q1D Fermi-surface structure. 
The large shoulder vanishes between $\varphi = 7 \pi/20$ and $9\pi/20$. 
The dimensional crossover between one and two dimensions 
appears only for $\varphi \lsim 3\pi/20$.

The FFLO critical field is low when $\varphi = \pi/2$. 
Considering the factors that suppress the FFLO state, 
particularly the orbital pair-breaking effect, 
this result suggests that FFLO modulation does not occur 
in directions $\varphi \approx \pi/2$ 
in the Q1D materials. 
Therefore, 
in the compound ${\rm (TMTSF)_2ClO_4}$, 
if the high-field phase for $\vH \parallel \vb'$ is an FFLO state, 
the modulation along $\vq$ and the vortices along $\vH$ 
cannot coexist in the form $\vq \parallel \vH$. 
In such a case, the Abrikosov functions with higher Landau-level indexes 
would contribute to the state, 
and the spatial modulation perpendicular to the magnetic field 
would be partly due to a paramagnetic effect~\cite{Shi97b,Shi09}.

\begin{figure}[htbp]
\vspace{2ex} 
\vspace{2ex} 
\begin{center}
\includegraphics[width=8.0cm]{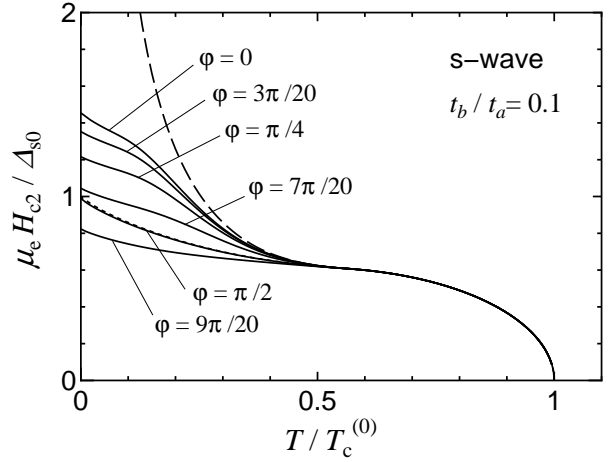}
\end{center}
\caption{
Temperature dependence of the upper critical field 
for s-wave pairing when $\tb/\ta = 0.1$. 
The solid curves show 
the results for the range of $\varphi$ from 0 to $\pi/2$. 
The dashed curve presents the result for a 1D system. 
The dotted curve is for a 2D isotropic system with s-wave pairing. 
}
\label{fig:hc2-T_various_phi_tb010_sw}
\end{figure}

Figure~\ref{fig:hc2-T_various_phi_tb010_sw} 
shows the results for s-wave pairing when $\tb/\ta = 0.1$. 
Similar to the result for d-wave pairing, 
a steep increase occurs just below $\Ttri$ for $\varphi \lsim 7\pi/20$, 
but the upturn at low temperatures occurs for all $\varphi$'s.

\begin{figure}[htbp]
\vspace{2ex} 
\vspace{2ex} 
\begin{center}
\includegraphics[width=8.0cm]{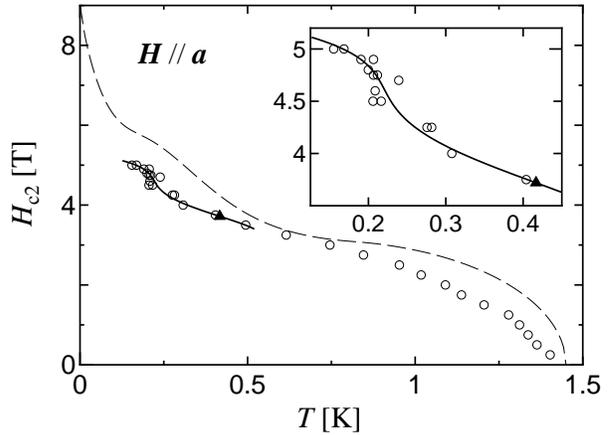}
\end{center}
\caption{
Comparison of the theoretical results and the experimental data 
for $\vH \parallel \va$~\cite{Yon08a,Yon08b}. 
The solid curve is a least-squares fit of the transition temperatures 
to a fourth-order polynomial in $H$ below $T = 0.5$~K. 
The thin dashed curve is the theoretical prediction for d-wave pairing 
with $\tb/\ta =0.1$, $\varphi = 0$, and $g/r = 1.5$, 
where $g$ and $r$ are the g-factor and a factor taking into account 
the correction of the ratio $\Delta_0/\Tc^{(0)}$, respectively. 
The closed triangle indicates the tricritical point obtained from 
the least-squares fit. 
}
\label{fig:H-T_fit4}
\end{figure}

In Fig.~\ref{fig:H-T_fit4}, the theoretical curve and experimental data 
are compared for $\vH \parallel \va$. 
The present theory neglects the strong coupling effect, 
impurity (disorder) pair breaking, and thermal fluctuations 
in the Q1D system. 
In particular, in the presence of these effects, 
the ratio $\Delta_0/\Tc^{(0)}$ would become larger 
than the weak coupling value of 1.76, 
where $\Delta_0$ denotes the superconducting gap at $T = 0$ and $H = 0$. 
The simplest way to take these effects into account 
is to change 
the ratio of the scaling of the $T/\Tc^{(0)}$ and $H/\Delta_0$ axes. 
Therefore, to compare the present result with the experimental data, 
we introduce the ratio $r = 
(\Delta_0^{\rm true}/\Tc^{\rm true})/(\Delta_{d0}/\Tc^{(0)})$, 
where $\Delta_0^{\rm true}$ and $\Tc^{\rm true}$ 
denote the true values of $\Delta_{0}$ and $\Tc^{(0)}$, respectively.

In the experimental data,~\cite{Yon08a,Yon08b} 
$\Hc2$ saturates near $T = 0.7~{\rm K}$ and $H = 3~{\rm T}$, 
reflecting the paramagnetic limit,\cite{Note2} 
but the increase in $\Hc2(T)$ recovers below $T \approx 0.3~{\rm K}$. 
Therefore, a tricritical point $(\Ttri,\Htri)$ 
should exist above $T = 0.3~{\rm K}$ 
if the recovery is due to the emergence of a different superconducting phase 
such as the FFLO state. 
The experimental data for $\Tc$ 
can be fitted by a fourth-order polynomial in $H$ 
over a region near and above $\Htri$, as shown in Fig.~\ref{fig:H-T_fit4}. 
As a result, a small shoulder appears below $T = 0.2~{\rm K}$, 
which is consistent with the theory. 
The point at which $\d^2 T(H)/\d H^2$ changes its sign 
is the tricritical point. 
The values determined by the least-squares fit are 
$\Ttri \approx 0.42~{\rm K}$ and $\Htri \approx 3.7~{\rm T}$. 
The difference between this $\Ttri$ and 
the theoretical $\Ttri = 0.56 \times \Tc \approx 0.81~{\rm K}$ 
is due to the orbital pair-breaking effect. 
In the present theory, 
$\Hc2(T)$ exhibits a second steep increase below the shoulder 
near $T = 0$ for d-wave pairing, while it does not for s-wave pairing. 
The complex behavior of $\Hc2(T)$ for d-wave pairing 
is due to the nesting effect in a Q1D system, 
reflecting the structure of the d-wave gap function with line nodes 
near $k_y = \pm \pi/2$. 
Therefore, if the orbital effect is not too strong, 
the second steep increase that indicates a d-wave FFLO state 
might be observed near $T = 0$ for $\vH \parallel \va$.

\section{\label{sec:last}
Summary and Discussion 
}

The temperature dependence of the upper critical field has been obtained 
for various ratios $\tb/\ta$ and directions of $\vq$ 
in strongly Pauli limited Q1D s-wave and d-wave superconductors. 
Several qualitatively different behaviors 
emerge depending on the parameters.

For $\tb/\ta \lsim 0.1$ and $\vq \parallel \va$, a new dimensional crossover 
from one dimension to two dimensions in $\Hc2(T)$ has been uncovered. 
The steep increase typical of 1D systems just below $\Ttri$ 
and the low-temperature upturn typical of 2D systems at low temperatures 
are connected by a shoulder at intermediate temperatures. 
This crossover occurs because the Fermi surfaces become diffuse owing to 
the thermal excitations at the energy scale $\kB T$. 
As argued in Sect.~\ref{sec:nesting}, 
the crossover temperature $T_0$ is proportional to $\tb \Tc^{(0)}/\ta$. 
For $T_0 \lsim T \lsim \Ttri$, 
the upper critical field behaves 
as if the Fermi-surface nesting is perfect. 
However, for $T \lsim T_0$ the Fermi surface becomes sharp enough to 
exhibit nesting on a line.

This unique temperature dependence for $\vH \parallel \va$ may be related to 
the nonmonotonic behavior of $\d \Hc2(T)/\d T$ that seems to be 
observed in ${\rm (TMTSF)_2ClO_4}$ at low temperatures. 
Detailed analysis of the temperature dependence 
that takes into account both the Fermi-surface effect 
and the orbital effect remains for a future study. 
If the orbital effect is too strong, 
the behavior of $\Hc2(T)$ would be simplified. 
If not, however, 
the second steep increase that indicates a d-wave FFLO state 
might be observed near $T = 0$.

For $\tb/\ta \gsim 0.15$, 
the temperature dependence of $\Hc2$ 
is qualitatively the same as that in Q2D systems, 
except for a slight shoulder at $0.2 \gsim \tb/\ta \gsim 0.15$. 
This behavior may correspond to the monotonic behavior of $\d H(T)/\d T$ 
in $({\rm TMTSF})_2 {\rm PF}_6$~\cite{Lee97} 
if the pairing is spin singlet.

The optimum FFLO modulation vector $\vq$ is parallel to 
the most conductive chain in the absence of the orbital effect, 
as confirmed by numerical calculations 
for both s-wave and d-wave pairings, independent of $\tb/\ta$. 
This direction of $\vq$ was previously assumed 
but is not {\it a priori} obvious 
when the Fermi surface is warped.

Curves for $\Hc2(T)$ have been presented 
when the magnitude of $\vq$ is optimized 
and the direction of $\vq$ is fixed in various directions. 
Unless the orbital pair-breaking effect is extremely weak, 
since $\vq \parallel \vH$,\cite{Gru66} 
the curves for $\Hc2(T)$ represent those for $\vH$ 
in the same direction as $\vq$, 
where the magnitude of $\Hc2$ is reduced by the orbital effect. 
Therefore, if one changes the direction of the magnetic field, 
the $T-H$ curve changes as shown in 
Figs.~\ref{fig:hc2-T_various_phi_tb010} and 
\ref{fig:hc2-T_various_phi_tb010_sw}, 
where the direction of $\vq$ is that of $\vH$.

For larger angles between $\vq$ and $\va$, 
the upper critical field is upward convex at low temperatures, 
and it converges to a finite value $\Hc2(0)$ with $\d \Hc2(0)/\d T = 0$, 
i.e., corresponding to type (c) in Table~\ref{table:nesting}, 
as for 3D systems. 
For the optimum $\vq$, the Fermi surfaces cross along lines, 
but they do not touch on a line. 
This situation is irregular, as explained in Sect.~\ref{sec:introduction}, 
and it originates from the small curvature of the Q1D Fermi surface. 
As a result, $\Hc2(T)$ exhibits 
a temperature dependence like that of a 3D system, 
but its magnitude is much larger than that in a 3D isotropic system.

When the orbital pair-breaking effect is extremely weak, 
the spatial variation of the gap function reflects 
that of the pure FFLO state, 
even when the angle between $\vH$ and $\vq$ is large 
(where $\vq$ is the optimum FFLO modulation vector 
for a vanishing orbital pair-breaking effect).
Then, the observed upper critical field is close to 
the maximum $\Hc2$ at each temperature, 
where $\vq$ is optimized in both magnitude and direction. 
In the present system, the maximum $\Hc2$ is 
that for $\vq \parallel \va$, 
as seen in Figs.~\ref{fig:hc2-T_various_phi_tb010} 
and \ref{fig:hc2-T_various_phi_tb010_sw}.

The present result that $\Hc2$ is largest when $\vq \parallel \va$ 
might appear to be inconsistent with the experimental result 
that the superconducting onset temperature is maximum 
when $\vH \parallel \vb'$~\cite{Yon08a} if $\vH \parallel \vq$. 
However, if the Fermi-surface effect is stronger than the orbital effect 
and the direction of $\vq$ is not strongly affected by the orbital effect, 
the result is consistent with the experimental result. 
Since the orbital pair-breaking effect is weakest 
when $\vH \perp \vq$~\cite{Cro12}, 
the present result $\vq \parallel \va$ implies 
that the superconducting onset temperature is maximum 
when $\vH \parallel \vb' \perp \va$. 
In this scenario, the Fermi-surface effect plays an essential role 
in locking the direction of $\vq$ to the direction of $\va$.

In conclusion, Q1D s-wave and d-wave superconductors 
with various values of $t_b/t_a$ and $\varphi$ 
exhibit qualitatively different behaviors of $\Hc2(T)$, 
including hybrid behaviors of types (a) through (c). 
In particular, when $\tb/\ta \lsim 0.1$ and $\varphi \lsim 3 \pi/20$, 
a new kind of dimensional crossover from one dimension to two dimensions 
has been discovered, which may be related to the behavior 
of the upper critical field for 
$\vH \parallel \va$ in $({\rm TMTSF})_2{\rm ClO}_4$.



\mbox{}

\noindent
{\bf Acknowledgments}

\vspace{0.5\baselineskip}

We would like to thank S. Yonezawa for helpful discussions and 
the experimental data. 



\end{document}